\documentclass[%
aps,%
prl,%
twocolumn,
superscriptaddress,%
showpacs,%
preprintnumbers,%
floatfix%
]{revtex4}

\usepackage{epsf}
\usepackage{ifthen}
\usepackage[usenames]{color}
\usepackage{amssymb,mathbbol,amsmath,amsbsy}
\usepackage{amsfonts}
\usepackage{dsfont}
\usepackage{amsmath}
\usepackage[dvips]{graphicx}
\usepackage[utf8]{inputenc}

\usepackage[dvips, colorlinks=true, bookmarks, pdftitle={},
pdfauthor={Haldun Sevincli}, pdfsubject={From LaTeX to PDF}, 
pdfkeywords={PDF, LaTeX, hyperlinks, hyperref}]{hyperref}

\newcommand{\atph}{\overline{\cal T}_\textrm{ph}}
\newcommand{\atel}{\overline{\cal T}_\textrm{el}}
\newcommand{\kph}{\kappa_\textrm{ph}}
\newcommand{\kel}{\kappa_\textrm{el}}
\newcommand{\lph}{\ell_\textrm{ph}}
\newcommand{\elph}{\overline{\ell}_\textrm{ph}}
\newcommand{\lel}{\ell_\textrm{el}}
\newcommand{\nph}{N_\textrm{ph}}

\begin{document}
\title{Enhanced Thermoelectric Figure of Merit in\\Edge Disordered Zigzag Graphene Nanoribbons}
\author{H. Sevin\c{c}li}
\author{G. Cuniberti}
\affiliation{%
Institute for Materials Science and Max Bergmann Center of Biomaterials,\\
Dresden University of Technology, D-01062 Dresden, Germany.}

\begin{abstract}
We investigate electron and  phonon transport through edge disordered zigzag graphene nanoribbons based on the same methodological tool of nonequilibrium Green functions.
We show that edge disorder dramatically reduces phonon thermal transport while being only weakly detrimental to electronic conduction.
The behavior of the electronic and phononic elastic mean free paths points to the possibility of realizing an electron-crystal coexisting with a phonon-glass.
The calculated thermoelectric figure of merit ($ZT$) values qualify zigzag graphene nanoribbons as a very promising material for thermoelectric applications.
\end{abstract}


\maketitle


The isolation of graphene \cite{novoselov:2004}, a one-atom thick $sp^2$-bonded planar carbon sheet, and quasi one-dimensional graphene nanoribbons (GNRs) have inspired research in many directions \cite{graphene:uygulama}.
GNRs are semiconductors with band gaps depending on their width \cite{gnr_tb,cohen:gnr_gap}.
The magnetic as well as electronic properties of GNRs are strongly dependent on edge shapes and very sensitive to width variations throughout the ribbon \cite{sevincli:gnr_superlattice}.
GNRs with average widths less than 10~nm are fabricated \cite{dai}, 
and control of GNR edges with an atomic precision has not been achieved yet, so edge disorder is an intrinsic property of GNRs.

Charge transport through edge disordered GNRs has been studied in detail in the last few years \cite{gnr_disorder:electronic,gnr_disorder:electronic_white}.
In particular, Areshkin \textit{et al.}~showed that GNRs with armchair edges are extremely sensitive to edge disorder whereas for those with zigzag edges (ZGNRs), the outstanding transport properties are weakly affected within the first conductance plateau (FCP), i.e.~the energy spectrum between the second conduction and valence band edges \cite{gnr_disorder:electronic_white}.

For thermoelectric energy conversion, on the other hand, a low thermal conductance is required and indeed high thermopower values up to  100~$\mu$V/K is reported for graphene \cite{graphene:tep}.
However two-dimensional graphene has extremely high thermal conductivity which is dominated by phonons \cite{graphene:thermal}, which nevertheless is strongly affected by graphene edges \cite{balandin:edge}.
The question is whether it is possible to overcome the high thermal conductivity so that GNRs might be good candidates for thermoelectricity.
Recently, significant reductions of phonon transport through nanotubes due to isotopic or Anderson-type disorder were reported \cite{bn_deney,mingo_isotopic_prl,derek_isotopic_nanolett,sevincli_cnt}.
It is also shown that phonon thermal conductance through Si-nanowires can be reduced by up to two orders of magnitude due to surface roughness or surface decoration which give rise to high thermoelectric coefficients \cite{si-nw}, as it was predicted theoretically that the thermoelectric figure of merit can be enhanced in low dimensional systems \cite{dresselhaus}.

In this Letter, we investigate electronic and phononic transport properties of edge disordered ZGNRs on an equal footing.
We show that $\kph$ can be reduced dramatically while the electronic transport can stay almost intact at the FCP, making edge disordered ZGNR an electron-crystal at these energies whereas a glass transition for phonons is achievable.

\begin{figure}[b]
        \begin{center}
        \includegraphics[scale=1]{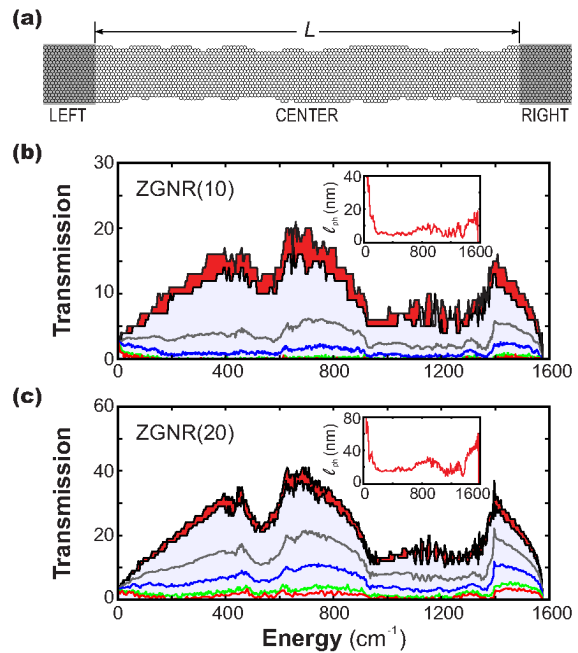}
	\end{center}
        \caption{
	(Color online)
	Schematics of the system (a). The central region with edge disorder has length $L$, and is connected to semi-infinite ZGNRs free of disorder.
	Transmission spectra of phonons, $\atph$, for pristine and disordered ZGNR are plotted for $N_z=$10 and 20 in (b) and (c), respectively.
	The dark (red) regions represent the difference in transmission between pristine ZGNR(10) and ZGNR(8) (b), and the same for ZGNR(20) and ZGNR(18) (c).
	$\atph$ is suppressed with increasing $L=$15.7, 63, 252 and 504~nm.
	Elastic mean free paths are given in the insets.
        }\label{fig:schema}
\end{figure}

\paragraph{Methods$-$}

We identify ZGNRs with the number of zigzag chains, $N_z$, which also determines their widths.
We use the common partitioning scheme for both electrons and phonons by dividing the system into three regions, namely left, right and central regions \cite{datta,cuniberti_kitap}.
The central region includes the disordered part, whereas the left and right regions are taken as semi-infinite perfect GNRs of given width (Fig.~\ref{fig:schema}).

For constructing the dynamical matrix, we use the fourth nearest neighbor force constant approximation (4NNFC), 
which yields phonon dispersions in agreement with density functional theory (DFT) calculations for graphene and carbon nanotubes \cite{saito,zimmermann,sevincli_cnt}.
For the case of GNRs, modification of force constants for the edge carbon atoms will improve the results \cite{hydrogenization,gnr_ribbon_hydrogen},
in the sense that it will result in a blue shift in the phonon density of states. 
Therefore a subsequent reduction of lattice thermal conductance and a further but minor improvement to our results in reducing lattice thermal conductivity can be expected, but it is neglected for the sake of simplicity.

The electronic part is modeled with a first nearest neighbor single orbital tight-binding (TB) Hamiltonian within the orthogonal parametrization  \cite{gnr_tb,gnr_tb_parameter}.
The TB Hamiltonian predicts a zero band gap for all ZGNRs while DFT based calculations show that all ZGNRs are semiconductors and their band gaps decrease monotonically with $N_z$, for $N_z > 4$ \cite{cohen:gnr_gap}.
For all $N_z$, highest valence band and the lowest conduction band give rise to a high density of states near the CNP ($E=0$) \cite{gnr_tb}.
The energy band gaps predicted by DFT calculations should not affect our conclusions because of the transport gap opening due to disorder.
In creating the disordered edges, we employ the algorithm explained in Ref. \cite{gnr_disorder:electronic_white}.
Edge atoms are eroded at each unit cell from both edges.
The number of carbon atoms to be removed ranges between 0 and 4 for each edge.
In calculating the transport properties, we follow an atomistic approach and employ nonequilibrium Green functions within the Landauer formalism \cite{landauer}, and a decimation technique \cite{decimation_detay} to obtain transmission amplitudes ${\cal T}_\textrm{ph}(\omega)$ and ${\cal T}_\textrm{el}(E)$ for phonons and electrons, respectively. 
We refer the reader to Refs. \cite{sevincli_cnt,cuniberti_kitap,datta} for details of the Green function technique.
We neglect the electron-phonon coupling in this work since it is shown that electron-phonon mean free path in ZGNRs is tens of $\mu$m at room temperature for ribbons having width $\sim$~10~nm \cite{white:e-ph}.
The thermoelectric figure of merit is defined as $ZT=S^2GT/\kappa$, where $S$ is thermopower, $G$ is electronic conductance, $T$ is temperature, and $\kappa=\kel+\kph$ is the thermal conductance with electronic and phononic contributions \cite{onsager,flensberg}.

\paragraph{Results and Discussions$-$}

We perform electron and phonon transport calculations for ZGNRs of two different widths, namely $N_z=10$ and $20$, with varying the length of the disordered region.
For each length $L$, an ensemble of edge disordered ZGNRs are generated.
After performing the electronic and phononic calculations for each configuration, we average the transmission spectra over the ensembles of 100 disordered edge profiles to obtain $\atel(E)$ and $\atph(\omega)$.
We quantify the relative amount of disorder, $\gamma_{N_z}$, as the average number of carbon atoms eroded from the edges of a unit cell of GNR divided by the number of carbon atoms in a pristine unit cell.
For the narrow ribbon, the relative amount of disorder is $\gamma_{10}=0.2$, while for the wider one it is $\gamma_{20}=0.1$.

\begin{figure}[t]
        \begin{center}
        \includegraphics[scale=1]{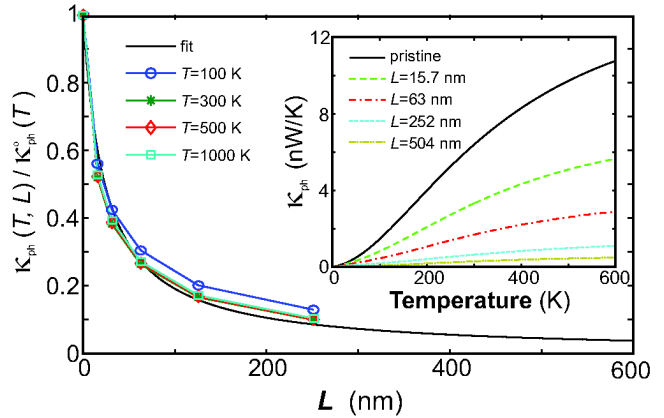}
	\end{center}
        \caption{
	(Color online)
        The ratio of phonon thermal conductance $\kph(T,L)$ to its pristine value $\kph^o(T)$ is plotted for ZGNR(20) at different temperatures. The ratio is fitted to a curve using Eq.~(\ref{eqn:fitting}). Inset shows $\kph$ of ZGNR(20) as a function of $T$.
        }\label{fig:phonon}
\end{figure}

Ensemble averaged transmission spectra of phonons, $\atph(\omega)$, are given in Fig.~\ref{fig:schema}(b) and (c) with varying lengths of the disordered region for $N_z=10$ and 20, respectively.
Pristine GNRs display staircase like transmission spectra, the transmission values corresponding to the number of available transport channels, $\nph$.
The transmission values drop significantly even for samples as short as 50~nm, the reduction is dramatic for longer samples.
Comparing the reduction of transmission values for two different widths having the same length, one observes that higher $\gamma_{N_z}$ values result in stronger suppression of phonon thermal transport.
One should note that different disorder types can give rise to differences in transmission reduction.
Our calculations show that edge disorder reduces phonon transport effectively for all energy values except very low energies.
On the other hand, isotopic disorder in carbon and BN nanotubes is shown to suppress highest energy modes more strongly than other modes \cite{bn_deney,mingo_isotopic_prl,derek_isotopic_nanolett}.
Likewise, Anderson-like disorder also distinguishes the high energy modes by suppressing them more effectively in carbon nanotubes \cite{sevincli_cnt}.
Using the relation 
$\atph(\omega)=\nph/(1+L/\lph)$ we calculate the elastic phonon mean free paths, $\lph$.
In the insets of Fig.~\ref{fig:schema}, $\lph$ are given as functions of energy for $N_z=10$ and 20.
Low energy modes preserve their quasi-ballistic behavior through edge disordered zigzag GNRs like they do under different disorder types \cite{murphy}.
For $\omega>50$~cm$^{-1}$, $\lph$ are quite short and always less than 75~nm for both $N_z$ values.
Such short mean free paths are observed only at high frequencies for isotopic or Anderson-like disorder.
The dramatic suppression of phonon transmission can be understood in the following way.
Anderson-like or isotopic disorder modifies the force constants which gives rise to elastic scatterings.
Edge disorder, on the other hand, not only modifies the vibrational frequencies but also changes the number of modes throughout the ribbon.
At an interface where the width of the ribbon changes, the number of transmission channels the ribbon can support is changed suddenly (Fig.~\ref{fig:schema}).
The difference in transmission coefficients is significantly large and distributed quite homogeneously at all energy values except $\omega\sim0$.
This is the reason why $\lph$ is of the same order of magnitude for all phonon spectrum except at $\omega\sim0$.
Phonon thermal conductance, $\kph(T)$, is suppressed strongly with increasing $L$ (Fig.~\ref{fig:phonon}).
Note that $\kph$ is already reduced by an order of magnitude for ZGNR(20) at room temperature when $L\sim250$~nm.
The length dependence of $\kph$ is stronger for the narrow ribbon, i.e.~for larger $\gamma$.
In Fig.~\ref{fig:phonon}, we plot the ratio of length dependent $\kph$ to its pristine value, $\kph^o$, at different temperatures.
We observe that the ratio follows a single curve for all lengths and at temperatures $T>50$~K.
At low temperatures, high frequency phonons are filtered out, and only the low frequency phonons, which have long $\lph$, conduct heat.
At higher temperatures this effect is less pronounced.
Since $\lph$ is oscillating around similar values except for $\omega\sim0$, a definition of an $\omega-$independent effective mean free path, $\elph$, is possible.
Letting $\atph(\omega)\simeq\nph(\omega)/(1+L/\elph)$, one can write
\begin{equation}
	\kph(T,L)\simeq\frac{\elph}{\elph+L}\kph^o(T).
	\label{eqn:fitting}
\end{equation}
This approximation accurately reproduces the numerical data for $L\gg\elph$, but it is not valid at temperatures $T<50$~K.
We perform a fitting and obtain $\elph=23.69~\textrm{nm}$ for ZGNR(20) and $7.04~\textrm{nm}$ for ZGNR(10). 
Note that the fit is in very good agreement with the computed data and we use the above relation for extrapolating $\kph(T,L)$ for $L>500~\textrm{nm}$.
For $L\gg\nph\elph$, complete localization of phonons in a very wide a range of the spectrum, and therefore a glass transition, is expected.


\begin{figure}[t]
        \begin{center}
        \includegraphics[scale=1]{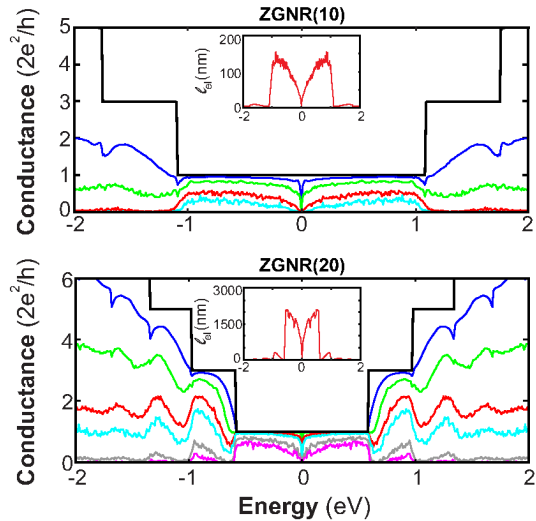}
        \caption{
	(Color online)
        Electron transport through edge disordered ZGNRs.
	Ensemble averaged transmission spectra $\atel$ are plotted for pristine ZGNRs and with varying sample lengths for $N_z=10$ (upper) and $N_z=20$ in (lower).
	Sample lengths are $L=$ 7.9, 31.5, 126, 252~nm (and additionally 1 and 2~$\mu$m for ZGNR(20)).
	Elastic mean free paths $\lel$ vs energy are extracted from the transmission data.
	Zero of the energy is set to the CNP.
        }
        \label{fig:electron}
        \end{center}
\end{figure}

Charge transport through edge disordered ZGNR shows interesting features.
In Fig.~\ref{fig:electron}, ensemble averaged transmission spectra of electrons $\overline{\cal T}_{el}(E)$ and calculated mean free paths, $\lel(E)$, are plotted for $N_z=10$ and 20.
The suppression of $\atel$ with increasing $L$ gives rise to different behaviors at different energies.
Once edge disorder is introduced, the transmission drops significantly at the CNP and a transport gap opens \cite{gnr_disorder:electronic_white,gnr_disorder:electronic}.
The width of the gap is determined by the relative amount of disorder as well as the length of the system.
the Opening of the transport gap is due to the fact that at the CNP the states are fully localized at the edges.
As $E$ is increased, the charge density is dispersed within the ribbon, and $\atel$ is very weakly affected by edge disorder.
It drops abruptly close to the band edges where the density of states is singular.
Outside the FCP, when $N_{el}>1$, the number of elastic scattering channels also increases and this results in a large reduction of transmission.
One should note that $\lel$ is reduced by approximately an order of magnitude when the amount of disorder is doubled.
With increasing sample length, a large derivative of $\atel$ is observed at the edges of the FCP and close to the CNP.
Such large derivatives are expected at energies where $\lel(E)$ changes abruptly, and we exploit this feature of edge disordered ZGNR for thermoelectric energy conversion.

\begin{figure}[t]
        \begin{center}
        \includegraphics[scale=1]{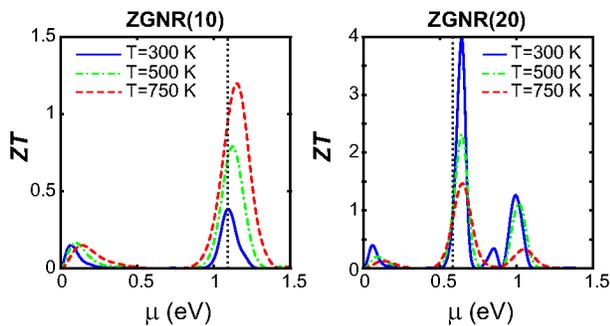}
	\end{center}
        \caption{(Color online) Thermoelectric figure of merit, $ZT$, versus chemical potential $\mu$ at different temperatures $T$ for ZGNR(10) (left) and ZGNR(20) (right) having lengths $L=0.25~\textrm{nm}$ and $4~\mu\textrm{m}$, respectively. Zero of $\mu$ is the CNP and the dotted lines represent the FCP edges.
	}
	\label{fig:zt}
\end{figure}

To achieve a high $ZT$, a low thermal conductance together with a high Seebeck coefficient (large derivative of $\atel$) and a high charge conductance are required \cite{onsager,flensberg}.
In ZGNRs, edge disorder suppresses $\kph$ by few orders of magnitude.
The derivative of $\atel$ is large at the CNP, but the transport gap is not sufficiently wide to overcome the mutual cancellation of electron and hole contributions to $S$.
Nevertheless, the cancellation decays at the edges of FCP with increasing $L$ as disorder suppresses charge transport  very strongly out of the FCP.
$ZT$ increases with increasing length for both ZGNRs, and the peaks appear close to the energies where the derivative of $\atel$ is large.
In Fig.~\ref{fig:zt}, $ZT$ is plotted at different temperatures as a function of the chemical potential $\mu$, and it is symmetric with respect to the CNP due to electron-hole symmetry.
The length of the sample is $L=0.25~\textrm{nm}$ ($L=4~\mu\textrm{m}$) for $N_z=10$ ($N_z=20$).
At room temperature $ZT$ reaches the values of 0.39 and 4 for ZGNR(10) and ZGNR(20), respectively.
For ZGNR(10) $ZT$ increases with temperature up to 1.2 at 750 K, whereas for the wider ribbon it decreases with temperature.
The inverse behaviors with temperature are due to the differences in the electronic transmission at energies beyond the FCP edges.
For ZGNR(10), $\atel$ is vanishingly small for $E>1.1~\textrm{eV}$, while for ZGNR(20) finite $\atel$ remains around $E=0.85~\textrm{eV}$, which also gives rise to the satellite peak of $ZT$ close to this energy.


In summary, we have shown that phonon thermal conductance can be suppressed significantly whereas charge transport stays intact within the first conduction plateau and a large derivative of the electronic transmission function is obtained at the edges of the plateau due to edge disorder in zigzag graphene nanoribbons.
High $ZT$ can be achieved depending on the ribbon width and the relative amount of disorder.
Furthermore, a sharp peak in the transmission spectrum, which maximizes $ZT$ \cite{mahan}, can be attained due to the fact that the extension of the first conduction plateau decreases with increasing ribbon width.

\paragraph{Acknowledgments$-$}
We acknowledge S. Roche for fruitful discussions.
This work was supported by the European Union project CARDEQ under contract No. IST-021285-2, 
and the priority program SPP-1386 of the German Research Foundation (DFG).
We further acknowledge the Center for Information Services and High Performance Computing (ZIH) at the Dresden University of Technology for computational resources.


\begin{thebibliography}{10}
\bibitem{hydrogenization}
TB Hamiltonian implicitly assumes edge passivation with hydrogen.
Theoretical calculations show that hydrogenated and dehydrogenated ribbons have very similar phonon dispersions \cite{gnr_ribbon_hydrogen}. 
Therefore we neglect the effects of hydrogenization in our phonon calculations.
\bibitem{onsager}
Onsager coefficients are calculated using the functions
$L_n(\mu,T)=\int dE\left(-\partial f(E,\mu,T)/\partial E\right)(E-\mu)^n\atel(E)$,
$f$ being the Fermi function.
Temperature dependent electrical conductance is $G(\mu,T)=(2e^2/h)L_0$.
Seebeck coefficient can be obtained as 
$S(\mu,T)=(-1/eT)L_1/L_0$, and the electronic contribution to thermal conductance is
$\kel(\mu,T)=(2/hT)(L_2-L_1^2/L_0)$.
\end{thebibliography}
\end{document}